\documentclass[
reprint,
groupedaddress,
aps,
pra,
]{revtex4-2}

\usepackage{graphicx}
\usepackage{dcolumn}
\usepackage{bm}

\usepackage{bbm} 
\usepackage{bm} 
\usepackage{dsfont} 
\usepackage{hyperref}
\usepackage{amsmath,amssymb,amsthm} 
\usepackage{mathtools} 
\usepackage{tensor} 
\usepackage{braket} 
\usepackage{ulem}
\usepackage{cancel}
\usepackage{ytableau}
    \ytableausetup{centertableaux}
    
\usepackage{tikz}
\usetikzlibrary{
    tikzmark,
    positioning, 
    decorations.pathreplacing,
    calligraphy,
    matrix
    }

\usepackage[dvipsnames]{xcolor}

\usepackage{hyperref}
\hypersetup{
	colorlinks=true, 
    linkcolor=[rgb]{0.91, 0.24, 0.43}, 
    citecolor=[rgb]{0.28, 0.24, 0.75}
    }

\allowdisplaybreaks 

\hypersetup{
    linkcolor=[rgb]{0.91, 0.24, 0.43}, 
    citecolor=[rgb]{0.28, 0.24, 0.75}
    }

\newcommand{\dd}{\text{d}}
\newcommand{\DD}{\mathrm{D}}
\newcommand{\pp}{\partial}

\renewcommand{\[}{\left[}
\renewcommand{\]}{\right]}
\renewcommand{\(}{\left(}
\renewcommand{\)}{\right)}
\renewcommand{\>}{\right\rangle}
\newcommand{\<}{\left\langle}

\renewcommand{\epsilon}{\varepsilon}
\newcommand{\spar}{\underline{\star}}
\newcommand{\ul}{\underline}
\newcommand{\zero}{\stackrel{!}{=}}

\begin{document}
{\phantom{.}\vspace{-2.5cm}\\\flushright Imperial-TP-KM-2026-01\\ }

\title{Linearised gravity as edge modes \& manifest duality and Lorentz covariance}%

\author{Calvin Y.-R. Chen}
\email{cyrchen@ntu.edu.tw}
\affiliation{Max Planck--IAS--NTU Center for Particle Physics, Cosmology and Geometry; \\
Leung Center for Cosmology and Particle Astrophysics; \\
Department of Physics and Center for Theoretical Physics, \\
National Taiwan University, Taipei 10617, Taiwan}

\author{Euihun Joung}%
\email{euihun.joung@khu.ac.kr}
\affiliation{Department of Physics, College of Science}
\affiliation{Research Institute for Basic Sciences,\\ 
Kyung Hee University, Seoul 02447, Republic of Korea}

\author{Karapet Mkrtchyan}
\email{k.mkrtchyan@imperial.ac.uk}
\affiliation{Abdus Salam Centre for Theoretical Physics, \\
Imperial College London, SW7 2AZ, United Kingdom}

\date{\today}

\begin{abstract}
We present the first formulation of linearised gravity in four dimensions which is manifestly Lorentz covariant and democratic, i.e. treats the two frames related by electric-magnetic duality on equal footing.
It is well-known that four-dimensional linearised gravity belongs to a class of singleton representations of the four-dimensional conformal algebra $\mathfrak{so}(2,4)$.
Our key insight is viewing this algebra as the isometry of $\text{AdS}_5$ and realising the massless spin-2 field as edge modes of a five-dimensional topological field taking values in a specific finite-dimensional representation of $\mathfrak{so}(2,4)$. 
The desired four-dimensional action is then found by a covariant boundary reduction procedure.
\end{abstract}

\maketitle



\section{Introduction}

Electric-magnetic duality is a fascinating symmetry of the eponymous electromagnetism, which rotates the electric and magnetic fields into each other in four spacetime dimensions \cite{Schrodinger:1935oqa,Deser:1976iy,Gaillard:1981rj, Deser:1981fr,Fradkin:1984ai, Sen:1992fr,Gibbons:1995cv}.
Linearised gravity is known to enjoy an analogous duality symmetry, which exchanges the linearised Riemann tensor and its dual.
The fact that this so-called electric-magnetic duality is an off-shell symmetry of the theory was shown explicitly in \cite{Henneaux:2004jw}. 

As opposed to electromagnetism and its $p$-form generalisations, for which Lorentz covariant approaches are available, the only action formulation of linearised gravity with manifest duality is the Hamiltonian formulation by Bunster and Henneaux \cite{Henneaux:2004jw,Bunster:2013oaa}.
This formulation does not retain manifest Lorentz covariance---the action is Lorentz invariant, but \textit{manifest} it is not. 
To date, no formulation exists which retains both manifest duality and Lorentz covariance---it is widely believed that these are in tension.
In this letter, we will derive an action formulation of linearised gravity in four spacetime dimensions in which invariance under Lorentz transformations and duality rotations are both manifest. 
This is made possible by first realising it as a boundary mode for a topological bulk theory.

Let us begin by recalling how electric-magnetic duality is realised in linearised gravity.
We consider four-dimensional Minkowski space $\mathbb{R}^{1,3}$, with coordinates denoted by $x^{\mu}$.
In the standard field-theoretic realisation, the fundamental degrees of freedom of linearised gravity are described by a rank-2 symmetric tensor $h_{\mu\nu}$ whose field strength is the linearised Riemann tensor 
\begin{equation}
    R^{\mu\nu}{}_{\rho\sigma} \equiv -2\pp^{[\mu}\pp_{[\rho}h_{\sigma]}{}^{\nu]}.
\end{equation}
By construction, this field strength satisfies 
\begin{equation}
    R_{\mu[\nu\rho\sigma]} = 0,\quad \pp_{[\lambda}R_{\mu\nu]\rho\sigma} = 0,
    \label{eq: lin spin 2 bianchi}
\end{equation}
which are known as the Bianchi identities. 
The dynamics of the helicity-2 field $h_{\mu\nu}$ in vacuum are then encoded in the Ricci-flatness condition (tracelessness of the field strength) 
\begin{equation}
    R^{\mu\nu}{}_{\mu\rho} = 0.
    \label{eq: lin spin 2 eom}
\end{equation}

Now, the totally anti-symmetric Levi-Civita tensor $\epsilon$ on $\mathbb{R}^{1,3}$ allows us to define the dual Riemann tensor as 
\begin{equation}
    \star R_{\mu\nu\rho\sigma} \equiv \frac{1}{2}\epsilon_{\mu\nu}{}^{\alpha\beta}R_{\alpha\beta\rho\sigma}.
\end{equation}
It is straightforward to see this also satisfies the Bianchi identities \eqref{eq: lin spin 2 bianchi} and the flatness condition \eqref{eq: lin spin 2 eom} on-shell.
Therefore, $\text{SO}(2)$ rotations of the Riemann tensor and its dual 
\begin{equation}
    \begin{pmatrix}
        R \\
        \star R
    \end{pmatrix}
    \mapsto
    \begin{pmatrix}
        \cos \theta & \sin \theta \\
        -\sin \theta & \cos \theta
    \end{pmatrix}
    \begin{pmatrix}
        R \\
        \star R
    \end{pmatrix}
\end{equation}
leave the dynamical content of the theory, i.e. \eqref{eq: lin spin 2 bianchi} and \eqref{eq: lin spin 2 eom}, invariant.
In \cite{Henneaux:2004jw}, it was in fact shown that there exist off-shell symmetry transformations that reduce to duality rotations of the Riemann tensor and its dual on-shell.
In other words, not only do duality rotations leave the equations of motion invariant, but they are a genuine symmetry of the theory, which are associated with a non-vanishing conserved charge generating these transformations \cite{Calkin:1965}.

Despite the fact that the linearised theory enjoys duality symmetry, the familiar formulation of full non-linear Einstein gravity clearly treats the fundamental field and its dual asymmetrically.
In this case, the choice of one over the other duality frame is justified by the fact that the duality symmetry of the free theory does not extend straightforwardly to the interactions \cite{Deser:2005sz,Henneaux:2017kbx, Monteiro:2023dev}.

In the linear theory (or at least when it is a good approximation \footnote{For instance, a prominent context in which linearised gravity describes the relevant dynamics at hand accurately is black hole perturbation theory. 
Here, electric-magnetic duality is related to Chandrasekhar duality and is responsible for isospectrality of the parity-even and -odd sectors \cite{Regge:1957td, Zerilli:1970wzz, Chandrasekhar:1975zza}.
This has significant consequence on physical observables such as the quasi-normal mode spectrum and Love numbers, see e.g. \cite{Solomon:2023ltn} for a review.
}), it is sensible to consider a formulation in which duality symmetry is manifest.
This is automatically achieved by any formulation in which the two dual field variables are treated on equal footing---such a formulation is said to be \textit{democratic}.
This is typically realised by doubling the field content and treating the field and its dual independently.
For linearised gravity, we relabel
\begin{equation}
    R^{(1)}_{\mu\nu\rho\sigma} \equiv R_{\mu\nu\rho\sigma},\quad R^{(2)}_{\mu\nu\rho\sigma} \equiv \star R_{\mu\nu\rho\sigma},
\end{equation}
and the original equations of motion can then be shown to be fully equivalent to 
\begin{equation}
    \star R^{(a)} = \epsilon^{a}{}_{b} R^{(b)},
    \label{eq: twisted self-duality}
\end{equation}
which is known as a \textit{twisted self-duality relation}.

The authors of \cite{Henneaux:2004jw,Bunster:2013oaa} (see also \cite{Barnich:2008ts, Henneaux:2016opm, Lekeu:2018kul, Henneaux:2019zod}) derive a democratic action for linearised gravity (in parallel with the formulation for spin-1 \cite{Henneaux:1988gg, Schwarz:1993vs, Deser:1997se, Henneaux:2020nxi}), by passing to the Hamiltonian formalism and solving the constraints in favour of so-called superpotentials which transform locally under off-shell duality rotations.
The price to pay is manifest Lorentz covariance.
These are analogous to the non-covariant formulations of chiral degrees of freedom in \cite{Floreanini:1987as, Henneaux:1988gg, Tseytlin:1990nb, Schwarz:1993vs}.

However, with an eye towards constructing more general classes of theories, retaining manifest Lorentz invariance is clearly at the very least as useful as making duality invariance manifest.
For that reason, in the case of $p$-forms, several approaches which recover manifest Lorentz covariance have emerged by now \cite{Pasti:1995ii, Pasti:1995tn, Pasti:1995us, Pasti:1996vs, Maznytsia:1998xw, Miao:2000fn, Sen:2015nph, Sen:2019qit, Mkrtchyan:2019opf, Avetisyan:2021heg, Evnin:2022kqn, Arvanitakis:2022bnr, Evnin:2023ypu, Hull:2023dgp}.
More recently, these have been systematically derived by a boundary reduction procedure of topological theories \cite{Arvanitakis:2022bnr,Evnin:2023ypu}.

The extension to spin-2 and higher is however non-trivial: The relevant gauge fields take values in a non-trivial representation of the spacetime symmetries, so that dualities mix with the latter.
This is an obvious obstruction to formulating such theories in a manifestly Lorentz covariant and democratic manner, and is precisely why no such formulation has been found to date.

In this letter, we will present a formulation of linearised gravity in four dimensions which preserves manifest Lorentz covariance and duality.
We thereby initiate a program which considerably extends the class of theories which we can describe democratically.
At the heart of our construction is the intimate connection of duality with conformal symmetry: Massless fields of spin-two (and higher) in four spacetime dimensions are so-called \textit{singleton} representations of the conformal algebra $\mathfrak{so}(2,4)$, which are special unitary irreducible representations of $\mathfrak{so}(2,4)$ which stay irreducible when restricted to its subalgebras $\mathfrak{iso}(1,3)$, $\mathfrak{so}(2,3)$, or $\mathfrak{so}(1,4)$.
This is despite the fact that the standard description in terms of the field potential does not have conformal symmetry \cite{Bracken:1982ny, Siegel:1988gd, Shaynkman:2004vu,Vasiliev:2007yc,Barnich:2015tma, Flores:2017ubo, Farnsworth:2021zgj}. 
To exploit this property of the singleton representations, we identify a starting point which manifests this $\text{SO}(2,4)$ conformal symmetry in four dimensions as $\text{AdS}_5$ isometry in the bulk. 
This enables us to construct a bulk topological theory which propagates four-dimensional linearised gravity as an edge mode, a possibility raised in \cite{Hinterbichler:2022agn}.
We are then able to build on the topological boundary reduction procedure developed in \cite{Arvanitakis:2022bnr,Evnin:2023ypu, Chen:2025xlo}.
More precisely, to arrive at the desired massless spin-two field theory on the boundary, we pick the anti-symmetric rank-three tensor representation of the conformal group with a specific set of asymptotic boundary conditions for the bulk topological theory.
Further generalisations to other singletons in higher dimensions and arbitrary spin will be spelled out in forthcoming work.

\section{Set-up \label{sec: set-up}}

Let us first set the stage by introducing the necessary ingredients for our set-up.

Consider $M=\text{AdS}_{5}$---we are ultimately interested in field theory on the conformal boundary $\pp M = \mathbb{R}^{1,3}$.
Let $x^{A}=(x^{\alpha},z)$ and $x^{\alpha}=(t,x^{i})$ be coordinates on $\text{AdS}_{5}$ and $\mathbb{R}^{1,3}$ respectively, such that the bulk metric in Poincar\'e coordinates takes the form
\begin{equation}
    \dd s^{2} = \frac{\eta_{\alpha\beta}\dd x^{\alpha}\dd x^{\beta}+ \dd z^{2}}{z^{2}}.
\end{equation}
It is also convenient to use embedding space $\mathbb{R}^{2,4} \supset \text{AdS}_{5}$, on which we will pick coordinates $x^{I} = (y,x^{A})$.
In that case, we will trade in the $z$- and $y$-coordinates for lightcone coordinates with $x^{\pm} = (z \pm y)/\sqrt{2}$ such that $x^{I} = (x^{-},x^{+},x^{\alpha})$.

Let us further consider a vector space $V$ of totally anti-symmetric 3-tensors of $\mathfrak{so}(2,4)$.
To distinguish elements of this vector space (the fibre) from anti-symmetric tensors on $M$ (the base) itself, we denote the components of the former using \underline{underlined} indices, e.g. 
for $A \in V$,
\begin{equation}
    A = \frac{1}{3!} A_{\ul{IJK}}V^{\ul{IJK}},
\end{equation}
where $V^{\ul{IJK}}$ are the generators of $V$.
Now, note that $V$ admits the invariant bilinear form
\begin{equation}
    \langle A, B\rangle = \frac{1}{3!} A_{\ul{IJK}}\,B^{\ul{IJK}}.
\end{equation}
Furthermore, the totally anti-symmetric Levi-Civita tensor $\epsilon$ of $\mathfrak{so}(2,4)$ defines an automorphism---the dualisation operator $\spar$, which acts as
\begin{equation}
    (\spar B)_{\ul{IJK}} = \frac{1}{3!} \epsilon_{\ul{IJK}}{}^{\ul{LMN}} B_{\ul{LMN}}.
\end{equation}
Since $\spar^{2} = -1$, this automorphism defines a complex structure, which allows us to decompose the vector space $V$ into two orthogonal vector spaces of anti-symmetric 2-tensors.
This split is not unique, but to make it explicit we can single out a direction by picking an $\text{SO}(2,4)$ vector $w_{\ul{I}}$.
Then, we can take these subspaces to be
\begin{equation}
    \begin{aligned}    
        &
        W^{(1)} \equiv \text{Span}_{\mathbb R}\left\{ W^{(1)\ul{JK}} \equiv w_{\ul{I}} \, V^{\ul{IJK}} \right\}, \\
        &
        W^{(2)} \equiv \text{Span}_{\mathbb R}\left\{ W^{(2)\ul{JK}} \equiv w_{\ul{I}}\, \(\spar  V\)^{\ul{IJK}} \right\}.
    \end{aligned}
    \label{eq: canonical splitting}
\end{equation}

Finally, we denote the $\mathfrak{so}(2,4)$-covariant derivative by $\DD$, which acts on tensors via the action of the commutator as $\DD \cdot = \dd \cdot + \[\Omega,\cdot\]$.
Here, $\dd$ is the standard exterior derivative on $\text{AdS}_{5}$ and $\Omega$ is the $\text{AdS}_5$ background connection taking value in the the symmetry algebra $\mathfrak{so}(2,4)$ of $\text{AdS}_{5}$.
This connection is flat, so the covariant derivative $\DD$ is nilpotent.
Explicit components of the connection and covariant derivatives acting on tensors are given in appendix \ref{app: covariant derivatives}.

\section{Democratic linearised gravity \label{sec: covariant boundary action}}

We are now in a position to present the theory which will provide a manifestly Lorentz covariant and democratic description of linearised gravity in four dimensions.

\subsection{Bulk description}

The theory of interest is a topological field theory on $M$, whose boundary degrees of freedom subject to certain constraints will eventually describe linearised gravity in a democratic manner.

Our starting point for this is the field $B$, which is a two-form in the base $\text{AdS}_{5}$ spacetime and takes values in the vector space $V$---it has indices in both the base $M$ and the fibre $V$.
Our theory of interest is described by the following action
\begin{equation}
    S = \int_{M} \langle \spar B \wedge \DD B\rangle + \frac{1}{2} \int_{\pp M} \langle B \wedge \star B\rangle 
    \label{eq: action 1}
\end{equation}
The bulk part of this is a Chern-Simons term with the covariant derivative $\DD$ with respect to the flat $\mathfrak{so}(2,4)$-connection.
The boundary term will be essential for what follows.

The classical equations of motion in the bulk and boundary are, respectively,
\begin{equation}
    \DD B =0,\quad \star B + \spar B = 0.
    \label{eq: bdry eom}
\end{equation}
The bulk field is closed and therefore pure-gauge, and the degrees of freedom in this theory hence reside purely within the boundary.
We further note that the boundary equation of motion already has the suggestive form of the twisted self-duality relation \eqref{eq: twisted self-duality} we are after.

On its own, however, \eqref{eq: bdry eom} propagates too many degrees of freedom, some of which are necessarily not unitary. 
To this end, we impose a specific set of asymptotic boundary conditions on the components of $B$ which differ for its various base and fibre components.
We claim this is precisely the necessary on-shell condition which leaves us with the correct degrees of freedom.
This is the higher-dimensional analogue of the conditions imposed in \cite{Coussaert:1995zp, Campoleoni:2010zq, Campoleoni:2011hg, Chen:2025xlo}.

\subsection{Boundary reduction}

Our goal is now to describe the boundary dynamics directly covariantly.
To that end, we perform the Lorentz covariant boundary reduction introduced in \cite{Arvanitakis:2022bnr}.
For this, instead of fixing a time direction (along which the usual Hamiltonian reduction is performed), we define a foliation of $\pp M$ (extended into $M$) by picking a nowhere-null closed one-form $v$.
Thanks to this, we can define the Lorentz covariant decomposition
\begin{equation}
    B =  C + v \wedge \tilde{R},
    \label{eq: reduction ansatz}
\end{equation}
for $C \in \Omega^{2}(M)$ with $\iota_{v}C = 0$, and $\tilde{R} \in \Omega^{1}(M)$, both taking values in $V$.
Substituting this into the bulk part of the action \eqref{eq: action 1}, we find that this yields
\begin{equation}
    \begin{aligned}
        S_{\text{bulk}} &= \int_{M} \< \spar C \wedge \DD C - 2 v \wedge \tilde{R} \wedge \spar \DD C \> \\
        &\hspace{2cm}+ \int_{\pp M}  \<v \wedge \tilde{R} \wedge \spar C\>.
    \end{aligned}
\end{equation}
In the bulk, $\tilde{R}$ acts as a Lagrange multiplier enforcing the condition that $v \wedge \DD C = 0$.
Since $\DD$ is nilpotent, this is locally solved by $C = \DD A + v \wedge P$, where $A,P \in \Omega^{1}(M)$ are $V$-valued \footnote{To show this, one can use arguments along the lines of \cite{Arvanitakis:2022bnr} and appendix C of \cite{Bansal:2021bis}.}. 
This amounts to taking our Ansatz \eqref{eq: reduction ansatz} to be
\begin{equation}
    B = H \equiv \DD A + v \wedge R,\quad R = \tilde{R} + P.
\end{equation}
We will frequently make use of the shorthand $F \equiv \DD A$.
When plugged back into the bulk, this Ansatz reduces it to a total-derivative term, and we are left with the following pure boundary action:
\begin{equation}
    S_{\text{bdry}}  =\frac12\int_{\pp M} \langle H \wedge\star H +2\,v\wedge R\wedge \spar F\rangle.
    \label{eq: action 2}
\end{equation}
As noted previously, this should be supplemented with a set of asymptotic fall-off conditions.
We will use the symbol $\zero$ to indicate that we are manually imposing the equivalence of the left- and right-hand side as a constraint.
As we shall see shortly, the following conditions turn out to pick out the correct four-dimensional theory:
\begin{equation}
    v\wedge F_{\ul{+\alpha\beta}} \zero \mathcal O(z^0),\quad v\wedge F_{\ul{-+\alpha}},\,v\wedge F_{\ul{\alpha\beta\gamma}}\zero \mathcal O(z),
    \label{eq: on shell conditions}
\end{equation}
where the leading power of $z$ of each field is determined by the conformal dimension of its fibre.
These can be imposed directly in the boundary action via Lagrange multipliers: 
\begin{equation}
    S_{\text{constr.}} = \int_{\pp M} v \wedge \chi \wedge F,
    \label{eq: action constraint}
\end{equation}
with $\chi^{\ul{+\alpha\beta}} = 0$.
In appendix \ref{app: constraints}, we show that $\chi$ is trivial on-shell.
Our theory is therefore specified by the total action
\begin{equation}
    S = S_{\text{bdry}} + S_{\text{constr.}}.
    \label{eq: total action}
\end{equation}
Subtleties that $S_{\text{constr.}}$ introduces to the forthcoming discussions regarding the dynamical content are postponed to appendix \ref{app: constraints}.

\subsection{Dynamical content \label{ssec: dynamical content}}

Let us begin by examining the boundary dynamics described by the action \eqref{eq: action 2} itself.

First note that the action \eqref{eq: action 2} has various gauge symmetries.
These, separately, comprise of 
\begin{itemize}
    \item $\DD$-exact shifts in $A$: $\delta A = \DD \alpha$,
    \item $v$-exact shifts in $R$: $\delta R = v\wedge \rho $,
    \item Combined shifts which leave $B$ invariant: $\delta A = v \lambda $, $\delta R = \DD \lambda$,
\end{itemize}
with $\alpha$, $\rho$, and $\lambda$ taking values in $V$.
This means that determining the physical content of the theory is not straightforward.

Let us make a few more comments on the boundary equation of motion.
From the equation for $R$, we see
\begin{equation}
	v \wedge \(\star H+ \spar H\) = 0.
    \label{eq: full boundary eom}
\end{equation}
Since $v^{2} \neq 0$, we can follow the steps outlined in \cite{Pasti:2012wv, Bansal:2021bis} to show that the expression within square brackets has to vanish:
\begin{equation}
    \star H + \spar H = 0.
    \label{eq: twisted self-duality H}
\end{equation}
The equation for $A$ is $\DD\(\star H-\spar H\)=0$, which, using \eqref{eq: twisted self-duality H}, simplifies to
\begin{equation}
    v \wedge \DD R=0.
    \label{eq: r pure-gauge}
\end{equation}
This means we can fix the gauge freedom in $R$ to rewrite \eqref{eq: twisted self-duality H} as
\begin{equation}
    \star \DD A + \spar \DD A = 0.
    \label{eq: pre twisted self-duality}
\end{equation}
as expected from the boundary equation of motion \eqref{eq: bdry eom}.
This is precisely the equation of motion we are after, i.e. a twisted self-duality relation of the form \eqref{eq: twisted self-duality}.
In appendix \ref{app: constraints}, we show the same conclusion holds when the constraints are accounted for in \eqref{eq: total action}.

\subsection{Lorentz covariance \label{ssec: lorentz covariance}}

For Lorentz covariance of the action defined by \eqref{eq: action 2} to be manifest, the one-form $v$ must not be fixed.
It therefore needs to be dynamical, with its equation of motion $\dd v=0$ being imposed by an additional Lagrange multiplier $\xi$ \cite{Evnin:2023ypu}:
\begin{equation}
    S_{v} = \int_{\pp M} \xi\wedge \dd v.
    \label{eq: action v}
\end{equation}
However, if this were to cause additional degrees of freedom to be propagated, the description of our theory would be spoiled.
Fortunately, the resulting theory has a gauge symmetry given by shifts of $v$ by exact forms, rendering $v$ pure-gauge. 
An equivalent perspective (closer in spirit to \cite{Pasti:1995tn, Bansal:2021bis, Evnin:2022kqn}) is that, taking $v=\dd a$ to be exact by Poincar\'e's Lemma (when $\xi$ goes on-shell), $a$ possesses a shift symmetry.
For the action \eqref{eq: action 2} alone, this follows directly in parallel to the spin-1 case \cite{Bansal:2021bis}. 
For the full action \eqref{eq: total action}, this argument requires suitable adjustments---this is described in appendix \ref{app: constraints}.

We are now ready to state our central result.
The full theory, defined by the action \eqref{eq: action 2} subject to the on-shell conditions \eqref{eq: on shell conditions}, is a manifestly Lorentz covariant and democratic formulation of linearised gravity in four dimensions.

\section{Non-covariant reduction \label{sec: non-covariant reduction}}

To make the connection to linearised gravity precise, we will now further reduce the action \eqref{eq: action 2} supplied with the asymptotic conditions \eqref{eq: on shell conditions} to the Bunster--Henneaux action derived in \cite{Henneaux:2004jw}. 
It is known that this is equivalent to the linearised Einstein--Hilbert action, so this effectively demonstrates that the degrees of freedom we are describing are indeed those of the free massless spin-two field.

To do so, we will now pick $v=\dd t$ and $w_{\ul{I}} =\delta_{\ul{I}}^{\ul{0}}$. 
This defines a foliation of the conformal boundary as $\mathbb{R}_{t} \times \Sigma_{t}$, where $\Sigma_{t}$ are spatial slices which are normal to $v$. 
We will see that with this choice our theory will reproduce their action after solving the constraints \eqref{eq: on shell conditions}.

\subsection{Gauge-fixing}

The gauge symmetries described in section \ref{ssec: dynamical content} are associated with zero modes of the action.
To proceed with the non-covariant reduction, our first step is therefore to fix the gauge by taking $\iota_{\pp_{t}} A = \iota_{\pp_{t}} R = 0$, such that only $A$ and $R$ with just spatial base components are non-zero. 
With explicit base indices (and implicit fibre indices), the action then takes the form
\begin{equation}
    \begin{aligned}
        S =& \int_{\pp M} \dd^{4}x\, \bigg<\frac{1}{2} \epsilon^{ijk} R_{i} \(\spar F\)_{jk} + \frac{1}{4} F_{\alpha\beta} F^{\alpha\beta} \\
        &\qquad \qquad +    R_{i} F^{0 i} - \frac{1}{2} R_{i} R^{i} \bigg>,
    \end{aligned}
\end{equation}
where we have taken $\epsilon_{0 ijk} \equiv \epsilon_{ijk}$.
The field $R$ appears quadratically and is hence auxiliary, so we may integrate it out.
Doing so, we are left with
\begin{equation}
    S = \frac{1}{2} \int_{\pp M} \dd^{4}x\, \bigg<F_{ij} F^{ij} + \epsilon^{ijk} F^{0}{}_{i}\(\spar F\)_{jk} \bigg>.
    \label{eq: action 3}
\end{equation}
Note the similarity of this with the Schwarz--Sen action \cite{Schwarz:1993vs} (\textit{modulo} the suppressed $\text{SO}(2,4)$ indices).

\subsection{Asymptotic conditions}

We will now solve the asymptotic conditions \eqref{eq: on shell conditions} to isolate the correct degrees of freedom.
As promised, this will reduce the action \eqref{eq: action 3} to the one in \cite{Henneaux:2004jw}.
For this, it is crucial to make the symmetries involving just spatial indices of both the base $M$ and the fibre $V$ explicit.
We do so by using the decomposition of the vector space $V =W^{(1)} \oplus W^{(2)}$ with respect to $w$ from \eqref{eq: canonical splitting}.
For a consistent boundary theory, it is also necessary to perform an asymptotic expansion of the fields in $z$, as defined in \eqref{eq: gauge fields}.
In the process of this, we also relabel various fibre components of $A$ by $\{Z,S,\omega,\phi\}$.

Now, at a practical level, the constraints \eqref{eq: on shell conditions} set the spatial components of $F_{\ul{\alpha\beta\gamma}}$, $F_{\ul{+\alpha\beta}}$, and $F_{\ul{-+\alpha}}$ (restricted to $\pp M$) to zero.
As an immediate consequence, we find that the only terms in the action that survive are
\begin{equation}
        S = -\frac{1}{2} \int_{\pp M} \dd^{4}x \,\epsilon_{ab}\, \epsilon^{ijk} \,F^{(a)}{}_{\ul{+m}0 i}\,F^{(b)}{}_{\ul{-}}{}^{\ul{m}}{}_{jk}\,.
    \label{eq: action 4}
\end{equation}

Before solving the asymptotic conditions further, let us first make our life easier by making use of the gauge redundancy $\delta A = \DD \Lambda$, for a gauge parameter $\Lambda \in \Omega^{0}(M)$.
The background connection $\Omega$ relates base and fibre indices, so we will now begin to see expressions which mix the two types of indices---this will eventually allow us to remove the distinction.
We denote the gauge parameters for the fields $\{Z,S,\omega,\phi\}$ by $\{\xi,\kappa,\rho,\sigma\}$ respectively (with the appropriate rescalings in $z$).
First note that the gauge transformation on $\phi^{(a)}$ takes the form
\begin{equation}
    \delta \phi^{(a)}_j = \pp_j \sigma^{(a)} + 4 \kappa^{(a)}_{\ul{j}}.
\end{equation}
It is clear that we can use the freedom in $\kappa^{(a)}$ to set $\phi^{(a)} = 0$ fully.
Similarly, a gauge transformation in $Z^{(a)}$ takes the form
\begin{equation}
    \delta Z^{(a)}{}_{\ul{i}j} = \pp_{j} \xi^{(a)}{}_{\ul{i}} - 2 \rho^{(a)}{}_{\ul{ji}} +2 \delta_{ij}\sigma^{(a)}.
\end{equation}
Here, we can use the freedom in our choice of $\rho^{(a)}$ (which is anti-symmetric) to fix $Z^{(a)}$ to be symmetric (it is now no longer necessary to distinguish its base and fibre indices).

Let us now return to the asymptotic conditions \eqref{eq: on shell conditions}.
In their most explicit form, these can be read off from the components of $F$ given in \eqref{eq: field strength components 3}.
We first consider the constraint on $F_{\ul{+ \alpha\beta}}$.
From \eqref{eq: torsionless constraint}, we see this allows us to solve for $\omega^{(a)}$ in terms of $Z^{(a)}$.
Then, since $Z^{(a)}$ is symmetric, the only non-zero projection of $\pp Z^{(a)}$ and hence $\omega$ is of $[2,1]$ Young-type symmetry.
By Bianchi's identity,
\begin{equation}
     \omega^{(a)\ul{jk}}{}_{i} = 2 \pp^{[j} Z^{(a)}{}_{i} {}^{k]}.
\end{equation}
Next, consider the asymptotic condition on $F_{\ul{+ - \alpha}}$, given by \eqref{eq: schouten algebraic}.
This is an algebraic constraint on $S^{(a)}$, forcing it to be symmetric (it is also now no longer necessary to distinguish its base and fibre indices).
Finally, we have the constraint on $F_{\ul{\alpha\beta\gamma}}$, given in \eqref{eq: schouten definition}. 
This equation has two sets of anti-symmetric spatial indices.
Since there is no rank-4, two-column Young diagram for $\text{SO}(3)$, the only content in this equation is therefore in its trace, and solving this algebraically gives
\begin{equation}
    \begin{aligned}
        S^{(a)}_{ij} &= - \frac{1}{4} \(\Delta Z^{(a)} - \pp_{m}\pp_{n}Z^{(a)mn}\)\delta_{ij}\\
	&\quad + \frac{1}{2}\(\pp_{i}\pp_{j}Z^{(a)} + \Delta Z^{(a)}_{ij}\) - \pp^{k}\pp_{(i}Z^{(a)}_{j)k},
        \end{aligned}
\end{equation}
where $\Delta \cdot \equiv \pp_{i}\pp^{i} \cdot$ and $Z \equiv Z^{i}{}_{i}$.
This fully solves the constraints \eqref{eq: on shell conditions} and allows us to write the unconstrained components of $F_{\ul{+\alpha\beta}}$ as
\begin{equation}
    \lim_{z\to0} z F^{(a)}{}_{\ul{+ i}0 j} =\pp_{0}Z^{(a)}{}_{ij} + \frac{1}{2} \epsilon_{i}{}^{mn} \epsilon^{a}{}_{b}\,\omega^{(b)}{}_{\ul{mn}j}.
\end{equation}
Plugging everything back into the action \eqref{eq: action 4}, we arrive at
\begin{equation}
    \begin{aligned}
        S &= - \int_{\pp M} \dd^{4}x\, \bigg[\epsilon^{ijk} \,\epsilon_{ab} \,\pp_{0}Z^{(a)}{}_{mi}\, \pp_{j} S^{(b)m}{}_{k} \\
		& \hspace{2.3cm} +  \frac{1}{2} \,\delta_{ab} \, S^{(a)\ell}{}_{i} \,\delta^{ijk}_{\ell mn}\,\pp_{j} \omega^{(b) \ul{mn}}{}_{k}\bigg],
    \end{aligned}
    \label{eq: action 5}
\end{equation}
after integration by parts.
It is straightforward to see that this is precisely the Bunster--Henneaux action presented in \cite{Henneaux:2004jw}, as promised.
The fields $Z^{(a)}$ precisely correspond to the so-called superpotentials in their formulation.
It is a democratic formulation: Manifest invariance under $\text{SO}(2)$ rotations between the $Z^{(a)}$'s is implied because $\epsilon$ and $\delta$ are invariant tensors of $\text{SO}(2)$.

\section{Concluding remarks}

In this letter, we presented the first formulation of linearised gravity in four dimensions with manifest Lorentz covariance and democracy.
To do so, we leverage the fact the free massless spin-2 fields in four dimensions belong to a class of singleton representations of $\mathfrak{so}(2,4)$.
In particular, we first use this insight to realise linearised gravity as the edges modes of a Chern-Simons theory for a two-form in $\text{AdS}_{5}$ valued in the anti-symmetric three-tensor representation of $\mathfrak{so}(2,4)$.
The boundary reduction procedure introduced in \cite{Arvanitakis:2022bnr} then leads to a manifestly Lorentz covariant boundary action \eqref{eq: action 2}.
Our central claim is then that, when supplied with the on-shell conditions \eqref{eq: on shell conditions}, the resulting boundary theory is a democratic description of linearised gravity in four dimensions.
We demonstrated this by breaking Lorentz covariance appropriately to recover the familiar democratic formulation of linearised gravity in four dimensions given by Bunster and Henneaux \cite{Henneaux:2004jw}.
Our work resolves this longstanding open problem and lends further credence to the idea that manifest Lorentz covariance and duality are not incompatible.

In higher dimensions, there are multi-form fields with rectangular Young diagram $[s^{p}]$, which are the natural generalisations of spin-$s$ singleton fields in $D=4$ to $D=2p+2$ spacetime dimensions.
This relies on the special property of how 
the rectangular $(p+1)$-row Young diagram representations of $\text{SO}(2,2p)$ split to two isomorphic $\text{SO}(1,2p)$ representations corresponding to rectangular diagrams with $p$ rows.
The case with generic $s$ and $p$ will be addressed in follow-up work.
For $p=2$, this describes massless higher-spin fields in four dimensions, of which a democratic description was given in \cite{Henneaux:2016zlu}.
For $s=2$ and $p=3$, these are conformal biforms featuring in so-called $(4,0)$-theory \cite{Hull:2000zn,Hull:2001iu,Henneaux:2017xsb} (extension to arbitrary $p$ is given in \cite{Joung:2012sa}).

The above corresponds to unitary conformal representations (singletons), but the pattern can be further generalised to more generic conformal representations, such as Fradkin--Tseytlin fields \cite{Fradkin:1985am}, Weyl-like fields of \cite{Joung:2012qy} (which will be shown to describe conformal representations in four dimensions), and generalisations to higher dimensions. 

\section*{Acknowledgements}

It is our pleasure to thank Arkady A. Tseytlin for feedback on an earlier draft  of this manuscript and helpful suggestions. 
We would also like to thank Marc Henneaux, Chris M. Hull, and Andrew J. Tolley for encouraging comments.

The research of CYRC was funded by Grant 115L104044 (Max Planck-IAS-NTU Center) supported by National Taiwan University under the framework of Higher Education SPROUT Project by the Ministry of Education in Taiwan.
The work of EJ is supported by the National Research Foundation of Korea (NRF) grant funded by the Korean government (MSIT) (No. 2022R1F1A1074977 and 2024-00466644). 
KM was supported by UKRI and STFC Consolidated Grants ST/T000791/1 and ST/X000575/1.
KM was also supported by the Higher Education and Science Committee of the Republic of Armenia, under the Remote Laboratory Program, grant number 24RL-1C047.

\bibliography{bibliography.bib}

\appendix

\section{Covariant derivatives \label{app: covariant derivatives}}

The generators $J_{\ul{MN}}$ of $\mathfrak{so}(2,4)$ satisfy
\begin{equation}
    [J_{\ul{MN}},J_{\ul{KL}}] = \eta_{\ul{MK}}J_{\ul{LN}} + \dots,
\end{equation}
where $\eta_{\ul{IJ}}$ is the flat metric on $\mathbb{R}^{2,4}$.
More precisely, one can show that the components of an arbitary element $A \in V$ are eigenvectors of $J_{\ul{-+}}$ with 
\begin{subequations}
    \begin{align}
        &[J_{\ul{+-}},A_{\ul{\alpha\beta\gamma}} ] = 0 \\
        &[J_{\ul{+-}},A_{\ul{\pm \alpha\beta}}] = \pm A_{\ul{\pm \alpha\beta}}\\
        &[J_{\ul{+-}},A_{\ul{+- \alpha}}] = 0,
    \end{align}
    \label{eq: bracket -+}%
\end{subequations}
and transform under $J_{\ul{\pm\alpha}}$ as
\begin{subequations}
    \begin{align}
        &[J_{\ul{\pm\lambda}},A_{\ul{\alpha\beta\gamma}} ] = 3\eta_{\ul{\lambda[\alpha|}}A_{\ul{\pm|\beta\gamma]}} \\
        &[J_{\ul{\pm\lambda}},A_{\ul{\mp \alpha\beta}}] = -A_{\ul{\lambda \alpha\beta}} \pm 2 \,\eta_{\ul{\lambda[\alpha|}}A_{\ul{-+|\beta]}} \\
        &[J_{\ul{\pm\lambda}},A_{\ul{\pm \alpha\beta}}] = 0 \\
        &[J_{\ul{\pm\lambda}},A_{\ul{-+ \alpha_{1} \dots \alpha}}] = A_{\ul{\pm \lambda \alpha}}.
     \end{align}
    \label{eq: bracket +a}
\end{subequations}

Next, note the non-zero components of the connection one-form $\Omega =\frac{1}{2} \Omega^{\ul{MN}}J_{\ul{MN}}$ are
\begin{equation}
    \Omega^{\ul{+-}} = \frac{\dd z}{z},\quad \Omega^{\ul{-\alpha}} = \sqrt{2} \frac{\dd x^{\alpha}}{z}.
\end{equation}

Putting everything together, the components of a component derivative $F \equiv \DD A$ can be shown to be
\begin{subequations}
    \begin{align}
        &\begin{aligned}
        F_{\ul{+ \alpha\beta}} =&  \dd A_{\ul{+ \alpha\beta}} +  z^{-1} \dd z \wedge  A_{\ul{+ \alpha\beta}} - \sqrt{2}z^{-1} \dd x^{\lambda}\wedge A_{\ul{\lambda \alpha\beta}} \\
        &- \sqrt{2}2\,z^{-1} \dd x_{[\alpha|} \wedge A_{\ul{-+|\beta]}}
        \end{aligned} \\
        &F_{\ul{\alpha\beta\gamma}} = \dd A_{\ul{\alpha\beta\gamma}} +3\sqrt{2}z^{-1} \dd x_{[\alpha|} \wedge A_{\ul{-|\beta\gamma]}} \\
        &F_{\ul{+- \alpha}} = \dd A_{\ul{+- \alpha}} - \sqrt{2} z^{-1}  \dd x^{\lambda} \wedge A_{\ul{- \lambda \alpha}} \\
        &F_{\ul{- \alpha\beta}} = \dd A_{\ul{- \alpha\beta}}  - z^{-1} \dd z \wedge  A_{\ul{- \alpha\beta}}.
    \end{align}
    \label{eq: field strength components}%
\end{subequations}

\subsection{Asymptotic expansions}

Consider the asymptotic expansions:
\begin{subequations}
    \begin{align}
        A_{\ul{+\alpha\beta}}(z,x^{\mu}) &= z^{-1} \[Z_{\ul{\alpha\beta}}(x^{\mu}) + \mathcal{O}(z)\], \\
        A_{\ul{\alpha\beta\gamma}}(z,x^{\mu}) &= \frac{1}{\sqrt{2}}\omega_{\ul{\alpha\beta\gamma}}(x^{\mu}) + \mathcal{O}(z), \\
        A_{\ul{-+\alpha}}(z,x^{\mu}) &= \frac{1}{\sqrt{2}}\phi_{\ul{\alpha}}(x^{\mu})+ \mathcal{O}(z), \\
        A_{\ul{-\alpha\beta}}(z,x^{\mu}) &= z^{+1} \[S_{\ul{\alpha\beta}}(x^{\mu}) + \mathcal{O}(z)\].
    \end{align}
    \label{eq: gauge fields}%
\end{subequations}
The choice of scaling in $z$ reflects the eigenvalues of the components of elements in $V$ under the dilatations given in \eqref{eq: bracket -+}.
With this, the leading-order terms in \eqref{eq: field strength components} become:
\begin{subequations}
    \begin{align}
        z^{+1} F_{\ul{+ \alpha\beta}} =& \, \dd  Z_{\ul{+ \alpha\beta}} - \dd x^{\lambda}\wedge \omega_{\ul{\lambda \alpha\beta}} - 2\dd x_{[\alpha} \wedge \phi_{\ul{\beta]}} + \mathcal{O}(z)\\
        F_{\ul{\alpha\beta\gamma}} =& \,\frac{1}{\sqrt{2}}\dd \omega_{\ul{\alpha\beta\gamma}} +3\sqrt{2} \dd x_{[\alpha} \wedge S_{\ul{\beta\gamma]}} + \mathcal{O}(z)\\
        F_{\ul{+- \alpha}} =& \,-\frac{1}{\sqrt{2}}\dd \phi_{\ul{\alpha}} - \sqrt{2}  \dd x^{\lambda} \wedge S_{\ul{\lambda \alpha}} + \mathcal{O}(z)\\
        z^{-1}F_{\ul{- \alpha\beta}} =&\,  \dd S_{\ul{\alpha\beta}} + \mathcal{O}(z).
    \end{align}
    \label{eq: field strength components 2}%
\end{subequations}
We can further write the fields in terms of the decomposition \eqref{eq: canonical splitting}.
Now making base indices explicit, \eqref{eq: field strength components 2} reads
\begin{subequations}
    \begin{align}
        \frac{z}{2} F^{(a)}{}_{\ul{+ i }}{}^{jk} &=  \pp^{[j} Z^{(a)}{}_{i}{}^{k]} - \omega^{(a)\ul{[j}}{}_{\ul{i}}{}^{k]} - \delta_{i}^{[j}\phi^{(a)\,k]} \label{eq: torsionless constraint} \\
        \frac{1}{\sqrt{2}}F^{(a)}{}_{\ul{+-}}{}^{ij} &= -\pp^{[i}\phi^{(a)\,j]} - 2 S^{(a)\ul{[i}j]} \label{eq: schouten algebraic}\\
       \frac{1}{\sqrt{2}} F^{(a)}{}_{\ul{ij}}{}^{mn} &= \pp^{[m} \omega^{(a)}{}_{\ul{ij}}{}^{n]} + 4 \delta_{[i}^{[m} S^{(a)}{}_{\ul{j]}}{}^{n]} \label{eq: schouten definition} \\
       \frac{1}{2z}F^{(a)}{}_{\ul{-i}}{}^{jk} &=\ \pp^{[j}S^{(a)}{}_{\ul{i}}{}^{k}.
    \end{align}
    \label{eq: field strength components 3}%
\end{subequations}

\section{Constraints \label{app: constraints}}

The theory of interest is described by the action \eqref{eq: action 2} subject to the on-shell conditions \eqref{eq: on shell conditions}.
Although there is \textit{a priori} no issue with implementing these manually, we may choose to implement these using a Lagrange multiplier $\chi$ as in \eqref{eq: action constraint}.
In this appendix, we will show how the discussions in sections \ref{ssec: dynamical content} and \ref{ssec: lorentz covariance} follow through regardless.

We begin by arguing the boundary equation of motion \eqref{eq: pre twisted self-duality} should continue to hold.
First note, we had to pick $\chi$ to be projected onto the subspace with $\chi^{\ul{+\alpha\beta}} = 0$.
Therefore in the subspace \textit{orthogonal} to $\chi$, the discussion in the main text applies straightforwardly.
The non-trivial part is the subspace \textit{parallel} to $\chi$.

Let us denote the components of $A$ parallel to $\chi$, i.e. for $A_{\ul{\alpha\beta\gamma}}$, $A_{\ul{-\alpha\beta}}$, and $A_{\ul{-+\alpha}}$, collectively as $A_{\parallel}$.
First note the equations of motion for $R$ are the same, i.e. \eqref{eq: twisted self-duality H} suitably projected.
Next, the equation for $\chi$ are precisely the constraints \eqref{eq: on shell conditions} themselves:
\begin{equation}
    v \wedge H_{\parallel} = 0.
\end{equation}
Because these are closed under fibre dualisation, the fibre dual of this should also hold.
Together with the equation of $R$, \eqref{eq: twisted self-duality H}, we then have
\begin{equation}
    v \wedge \spar H_{\parallel}=0 \quad \implies \quad v \wedge \star H_{\parallel} = 0.
\end{equation}
For the same reasons as mentioned below \eqref{eq: full boundary eom}, this, combined with the original form of the constraint, implies $H_{\parallel} = 0$ itself.
From this, the projection of \eqref{eq: r pure-gauge} onto the subspace parallel to $\chi$ follows, and we infer $R_{\parallel}$ is pure-gauge.
Now, the equation for $A_{\parallel}$ is
\begin{equation}
    \DD\(\star H_{\parallel} - \spar H_{\parallel} + v\wedge \chi\)=0.
\end{equation}
However, we just showed that $H_{\parallel}=0$, so 
\begin{equation}
    v\wedge \DD\chi=0.
    \label{eq: chi constraint}
\end{equation}
Now, note that the action \eqref{eq: total action} is also invariant under gauge transformations $\delta \chi=\DD \alpha + v\wedge \beta$. 
Therefore, just as for $R$, \eqref{eq: chi constraint} implies that $\chi$ is pure-gauge.

The observation that $H_{\parallel} = 0$ also enables to argue that the one-form $v=\dd a$ has a shift-symmetry by exact forms.
For dynamical $v$, the equation of motion for $a$ is
\begin{equation}
    \dd\left\langle R \wedge (\star H + \spar H) - \chi \wedge H_{\parallel}\right\rangle =0.
\end{equation}
This vanishes identically as a consequence of the other equations of motion, implying $a$ has a shift symmetry.

\end{document}